\documentclass[aps,floatfix,pra,superscriptaddress,twocolumn,nofootinbib]{revtex4}

\usepackage{graphicx}
\usepackage{graphics}
\usepackage{amssymb}
\usepackage{amsmath}
\usepackage{latexsym}
\usepackage{epstopdf}
\newcommand{\ket}[1]{| #1\rangle}
\newcommand{\bra}[1]{\langle #1 |}

\newcommand{\sx}{\hat{\sigma}_{x}}

\begin{document}
\title{Decoherence in quantum walks and quantum computers}
\author{Andrew P. Hines}
\address{Pacific Institute of Theoretical Physics,Department of Physics and Astronomy, University of British Columbia,6224 Agricultural Rd, Vancouver BC, Canada V6T 1Z1}
\author{P.C.E. Stamp}
\address{Pacific Institute of Theoretical Physics,Department of Physics and Astronomy, University of British Columbia,6224 Agricultural Rd, Vancouver BC, Canada V6T 1Z1}
\begin{abstract}
Decoherence is the major stumbling block in the realization of a
large-scale quantum computer. Ingenious methods have been devised to
overcome decoherence, but their success has been proven only for
over-simplified models of system-environment interaction. Whether
such methods will be reliable in the face of more realistic models
is a fundamental open question. In this partly pedagogical article,
we study two toy models of quantum information processing, using the
language of \emph{quantum walks}. Decoherence is incorporated in 3
ways - by coupling to a noisy `projective measurement' system, and
by coupling to oscillator and spin baths.
\end{abstract}

\maketitle

\section{Introduction}

Any realistic quantum computer must be implemented
\emph{fault-tolerantly} -- the ideal quantum computation must be
reliably performed using non-ideal components \cite{NC00}. The great
difficulty, when compared with classical information processing is
that the ``errors'' that can affect qubits are much more
complicated. Classical bits can only be in one of two states (`0'
or `1'), so the sole possible error is an unwanted change of the
bit-value -- a bit-flip. In the case of quantum bits, there is a
continuum of different errors that may occur.  Qubits can be placed
in arbitrary superposition $\alpha |0\rangle + \beta |1\rangle$ and
with multiple qubits, superpositions give entangled states.
Superpositions  can \emph{decohere} into statistical mixtures of
states, destroying the phase correlations necessary for a quantum
computation.

Decoherence is caused by interactions between a quantum system and
its environment \cite{Wei99,PCES06}. A system starting in some
superposition of states becomes entangled with its environment, and
the resulting state cannot be decomposed into a simple product
state. Averaging over the environmental variables gives a full, or
partial, statistical mixture of states, described by the reduced
density matrix of the system, rather than a coherent superposition.

To be able to perform quantum computation in the face of
decoherence, some form of quantum error correction is necessary. One
of the major breakthroughs in the field of quantum information
processing was the discovery of quantum error correcting protocols
\cite{Sho95,Ste96,CS96}. The continuum of potential single-qubit
errors can be overcome by correcting for just a discrete subset
\cite{NC00}, by encoding a single logical qubit in a number of
entangled qubits. It was then claimed that if the total strength of
the decoherence is below a certain, finite threshold value, any
quantum computation could be performed to some arbitrarily close
accuracy -- this is a \emph{fault-tolerant threshold}. Estimates of
threshold values depend greatly upon the quantum error-correcting
code used, assumptions about the architecture of the quantum
computer, and most importantly, how the interaction with the
environment is described.

The simplest approach, used in almost all papers until recently, was to
treat the environment essentially as a stochastic noise source --
decoherence was modelled as a sequence of independent errors, each
affecting a single qubit, occurring randomly in time. Each ideal
unitary component of a quantum computation can then be replaced by a
non-unitary map describing evolution with some probability of an
error. Such noise models were used to derive fault-tolerant
thresholds bounding the probability of an error occurring at each
location in the circuit \cite{AO99,Kni04}. Two key assumptions about
the interactions with the environment are implicit in these models:
they are local -- meaning no correlations between different qubits,
except through gates -- and Markovian -- the environment is
memory-less, so there are no correlations between different times.

Neither of these assumptions is realistic; they miss all the
non-local effects (in space and time) which result when a set of
quantum systems are coupled to a real environment. Quantum
environments are usually described by either oscillator baths
\cite{Wei99,CL83}, or spin-baths \cite{PCES06,PS00} -- representing
delocalized or localized environmental modes, respectively -- or a
combination of both. Applying such models to a fault-tolerance
calculation requires a Hamiltonian formulation, starting with a
description of the system and environment dynamics, and their
interaction.

Recently, several attempts were made to introduce more realistic
assumptions. These include analysis of a local, but non-Markovian,
environment \cite{BT05}, then extended to include long-range spatial
correlations between qubit errors \cite{AGP06,APK06}. In these
analyses the threshold value was derived in terms of an operator
norm on the interaction term in the full system-environment
Hamiltonian - this operator norm depended explicitly on an arbitrary
high-energy cut-off in the Hamiltonian. This makes no physical sense
-  no physical result should depend on the value chosen for these
cut-offs. Clearly one needs a formulation of the problem which is
based on the physical mechanisms operating in the system. Certainly
this is possible - indeed, several quantitative analyses of
correlated errors for real systems exist already in the
literature \cite{bar06,bar07,MST06}, including one which incorporated
long-range dipolar interactions \cite{MST06}.

This paper discusses one way of setting up such a formulation
(although there is no space here for a complete discussion). We use
two `toy models' discussed below -- our goal is frankly to give some
intuition for the problem. Our approach is formulated in the
language of \emph{quantum walks}. A quantum walk describes the
dynamics of a particle on some arbitrary mathematical graph, in
general coupled to an external environment. Hamiltonians describing
a `quantum walker' can be mapped to a very large class of
Hamiltonians describing quantum information processing systems.
Again, we shall see that a realistic formulation of environmental
couplings in these Hamiltonians is necessary to get sensible
results.

We begin with a brief discussion of quantum walks without
environments, showing how to reformulate these in a Hamiltonian
framework, and to map between quantum walk Hamiltonians, spin
networks and quantum gates. We then extend these `free walker'
models to general models of walks coupled to quantum environments.
We then look at our two toy models of quantum walks (with the walker
on a hypercube and hyperlattice respectively). We compare how the
walker dynamics is modified using 3 models of decoherence -- first a
diagonal coupling to a Markovian `noise' environment, then a
diagonal coupling to an oscillator bath, and finally a non-diagonal
coupling to a spin-bath. No attempt is made to give a complete
treatment - our aim is to show, in these simple models, the kind of
results one gets by using physically realistic environments to
analyze decoherence (and how radically different these are from the
models usually used in error-correction theory).

\section{`Bare' Quantum Walks}

Quantum walks were introduced as the quantized version of classical
random walks \cite{ADZ93}, which are one of the cornerstones of
computer science \cite{Pap94,Kem03}. They  display fascinating
behaviour quite different from their classical counterparts
\cite{Kem03,FG98,childs02,AKR05,kempe02}. One motivation for
introducing them was the hope that one could thereby generate new
kinds of quantum algorithm, which have so far proved very hard to
find. The quantum walk algorithms proposed so far fall into one of
two classes \cite{ambainis03}. The first is based on exponentially
faster hitting times \cite{FG98,childs02,kempe02,childs02a}:
generally the hitting time is defined as the mean `first passage'
time taken to reach a given target node from some initial state. The
second class provides a quadratic speed-up using a quantum walk
search \cite{childs04,shenvi03}. In the case of a spatial search,
the quantum walk algorithms can outperform the usual quantum
searches based on Grover's algorithm.

`Bare' quantum walks (ie., with no environment) are defined over an
undirected graph with $N$ nodes, each labelled by an integer $n\in
[0,N-1]$. Suppose we now want to describe them in a Hamiltonian
framework. We can then define `\emph{simple} quantum walks'
\cite{HS07} by the Hamiltonian
\begin{equation} \label{Ham-qwa}
\hat{H}_o^s = -\sum_{ij}\left(\Delta_{ij}(t)\hat{c}_{i}^{\dagger}
\hat{c}_j + H.c.\right) \;+\; \sum_j \epsilon_j(t)
\hat{c}_{j}^{\dagger}\hat{c}_j.
\end{equation}
Here a walker at node $n$ corresponds to the quantum state $\ket{n}
=\hat{c}_n^{\dagger}\ket{0}$. The first term in (\ref{Ham-qwa}) is a
`hopping' term with amplitude $\Delta_{ij}(t)$ along the link
$\{ij\}$ between nodes $i$ and $j$; the second describes `on-site'
node energies $\epsilon_j(t)$. This is a generalization of the
continuous-time quantum walks originally introduced by Farhi and
Gutmann \cite{FG98}. One can also imagine a walker coupled to
some other degrees of freedom, which partially control the walker
dynamics (an example of this is where the walk is controlled by a
discrete `coin' variable). We therefore define a `\emph{composite}
quantum walk' Hamiltonian, in which the simple walker couples at
each node $j$ to a mode with Hilbert space dimension $l_j$, and on
each link $\{ij\}$ to a mode with Hilbert space dimension $m_{ij}$,
giving:
\begin{eqnarray}
 \label{Ham-qwb}
\hat{H}_o^C &=& -\sum_{ij} \left(F_{ij}({\cal
M}_{ij};t)\hat{c}_{i}^{\dagger}\hat{c}_j  +H.c. \right) \nonumber \\
&& + \sum_j G_j({\cal L}_j;t)\hat{c}_{j}^{\dagger}\hat{c}_j \;+\;
\hat{H}_o(\{{\cal M}_{ij}, {\cal L}_j \}).
\end{eqnarray}
While there have been proposals for implementing quantum walks in
real space \cite{TM02,DRKB02}, it is more likely that the walk will
take place in the `information' space of some other physical system,
like a register of qubits.

Within this general formulation, one can map between qubit
Hamiltonians, spin Hamiltonians, quantum gate systems, and quantum
walk Hamiltonians \cite{HS07}.  In Ref. \cite{HS07} we construct
such mappings, exhibiting different ways of encoding quantum walks
in multi-spin/qubit systems. These mappings are useful to explore
quantum algorithms and quantum information processing hierarchies.
On the more practical side, they offer a general way of implementing
quantum walks over various graphs. In this paper we will use them to
study the dynamics of quantum information processing, particularly
the dynamics of decoherence in the presence of realistic quantum
environments.

\begin{figure}
\centering \scalebox{0.6}{\includegraphics{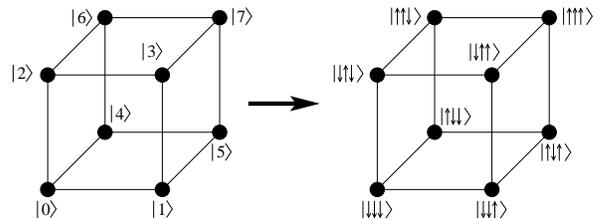}}
\caption{Three qubit encoding for the quantum walk on the cube,which can be easily generalized to higher dimensions using
additional qubits. Each transition between nodes corresponds to the
bit-flip of a single qubit.}
 \label{fig::cubic-lattice}
\end{figure}

As examples, consider the two toy models we will analyze. The first
is a symmetric hypercube -- in the Hamiltonian (\ref{Ham-qwa}), one
assumes constant nearest-neighbour hopping between nodal sites (ie.,
$\Delta_{ij} = \Delta_o, \,\,\forall \; i,j$), and no on-site terms
(ie., $\{ \epsilon_j \} = 0, \;\; \forall \; j$), in a network
restricted to a single $D$-dimensional hypercube. This is equivalent
to the simple qubit Hamiltonian
\begin{equation}
H_o^{HC} = -\Delta_o \sum_{n=1}^D \hat{\tau}^x_n,
 \label{Do}
\end{equation}
describing $D$ independent qubits, with the Pauli matrix $\hat{\tau}_j$
describing the $j$-th qubit. The node states for the walk are
encoded in terms of multi-qubit states, using the binary
representation of the node number, as shown in figure
\ref{fig::cubic-lattice}.

Our second toy model extends this $D$-dimensional hypercube to an
infinite $d$-dimensional cubic hyperlattice, so the bare Hamiltonian
is now
\begin{equation}
\hat{H}_o^{HL} = -\Delta_{o}\sum_{ij}
(\hat{c}_{i}^{\dagger}\hat{c}_j + H.c.) \;\; \equiv \;\; \sum_{\bf
p} \epsilon_o({\bf p}) \hat{c}_{\bf p}^{\dagger}\hat{c}_{\bf p}
 \label{H-HLo}
\end{equation}
where ${\bf p}$ is a `quasi-momentum' (called the `crystal momentum'
in the solid-state literature); and the `band energy' is
\begin{equation}
\epsilon_o({\bf p}) = 2\Delta_o \sum_{\mu = 1}^d \cos (p_{\mu}a_o)
 \label{Eop}
\end{equation}
where we assumed a cubic lattice spacing $a_o$; henceforth we put
$a_o = 1$.

\section{Decoherence in Quantum Walks}

The effects of decoherence on quantum walks have been considered by
several authors, mostly for discrete-time quantum walks (for a
review see \cite{Ken06}). Most investigations to date have either
used simple, Markovian `noise' sources
\cite{KT03,BCA03a,BCA03b,BCA03c,Fed06,AR05}, or imperfect
(non-unitary) evolution \cite{MBSS02}, to produce decoherence. In
the discrete-time walk, the effect of the environment is represented
as a weak measurement on the system: at each time step, there is
some probability $p$ that a measurement is made, and the outcome
lost to the environment. This could affect the `coin' and/or walker
degree of freedom. The measurement monitors the position of the
walker and state of the `coin', in the basis coupling to the walker
transition, though other mechanisms on the coin have been
considered. Extensions to continuous-time walks have similarly
described the environment as a position-monitoring measurement
device \cite{Fed06}.

The focus of such models has primarily been to demonstrate the
transition of the quantum walk to a classical random walk
\cite{BCA03a}, even for quite exotic models \cite{BCA03c}. The
analysis has used random measurements to simulate decoherence. Using
such analyses it is argued that decoherence may even be
\emph{useful} in some quantum walks \cite{KT03}, and that one gets a
quantum speedup of classical mixing processes \cite{Ric06}. However
it has also been argued that quantum walks {\it must} approach a
classical random walk in the long-time limit when decohered
\cite{Ken06}.

In this paper we use a Hamiltonian description which allows us to
incorporate realistic couplings to the kinds of environment which
exist in Nature. It then becomes possible to solve for the dynamics
of quantum walk systems, without using ad hoc decoherence models.
This approach gives very different results from those noted above
(for example, one sees in some cases quite dramatic departures from
classical random walks in the long-time limit). It also has the
great advantage of bringing work on decoherence in both quantum
walks and qubit networks into contact with experiment.

Although there is a large variety of different coupling mechanisms
in Nature, one can classify them in a simple way. First, the
environments can be divided into two classes. Environments of {\it
extended} or delocalised modes (like phonons, photons, conduction
electrons, spin waves, etc) can be mapped to an oscillator
bath\cite{Wei99}, whereas environments of localised modes (local
spins, including nuclear spins, two-level systems, defects,
dislocations, etc.) must be mapped to a spin bath \cite{PS00}; the
two environments have very different effects on some central quantum
system coupled to them \cite{PCES06}. The second classification one
can make is between diagonal and non-diagonal environmental
couplings to the central quantum system. The former couple to
individual states of the central system system, the latter to
transitions between these states (thus the distinction depends on
which basis we choose). In the case of quantum walks a diagonal
coupling has the form $H_{int} = \sum_{j \alpha} U_j(X_{\alpha})
n_j$, where $n_j = \hat{c}_j^{\dagger} \hat{c}_j$, and the $\{
X_{\alpha} \}$ are the environmental modes, which couple here to
states localised on a node. A non-diagonal coupling couples the $\{
X_{\alpha} \}$ to inter-node transitions, and has the form $H_{int}
= \sum_{ij, \alpha} V_{ij}(X_{\alpha}) [\hat{c}_i^{\dagger}
\hat{c}_j + H.c.]$. We will see examples of these various classes of
coupling in what follows.

We now turn to the dynamics of decoherence in our two toy models.

\subsection{Quantum walk on the hypercube}

For a simple hypercube the free walker dynamics is trivial. Thus,
for a walker initialized at the $\vec{z} = 0 \equiv \downarrow
\downarrow \ldots \downarrow$ corner, the probability of being at an
arbitrary site $\vec{z}$, as a function of time, is
\begin{equation}
P_{\vec{z} 0 }(t) = \cos^{2n_{\downarrow}}(2\Delta_o
t)\sin^{2n_{\uparrow}} (2\Delta_o t),
 \label{Pz}
\end{equation}
where $n_{\downarrow}$ is the number of $\downarrow$'s, and
$n_\uparrow$ the number of $\uparrow$'s appearing in $\vec{z}$.

\subsubsection{Master equation approach}

We begin by simply coupling to a memory-free system which randomly
monitors the walker `position' on the hypercube. One then describes
the dynamics by the master equation
\begin{equation}\label{eq::master}
\frac{d\rho(t)}{dt} = -i\left[\hat{H},\rho(t)\right] + \gamma
\sum_{k=1}^{N}\mathcal{D}\left[ \hat{M}_k \right] \rho(t),
\end{equation}
where
\begin{equation}
\mathcal{D}\left[ \hat{M}_k\right] \rho= \hat{M}_k \rho
\hat{M}_k^{\dagger} - \frac{1}{2}\left( \hat{M}_k^{\dagger}
\hat{M}_k \rho+ \rho \hat{M}_k^{\dagger} \hat{M}_k  \right),
\end{equation}
such that the  $\{\hat{M}_k \}$ describe the interaction with the
environment; this is the continuous-time generalisation of a
discrete-time description\cite{Ken06}.  There are two obvious ways
to couple to the walker position. The usual approach \cite{Ken06}
has been to have $\{ \hat{M}_k \}$ project onto individual hypercube
nodes, ie., $M_k  \rightarrow |j\rangle \! \langle j |$. In a qubit
representation, this site-based decoherence actually corresponds to
a rather unphysical multi-qubit interaction with the environment. A recent paper\cite{AR05} claims to consider
this standard case, however, they actually choose projectors onto the
single qubit basis states for the $M_k$'s. This is clearly a more
physical representation of the decoherence when the quantum walk is implement on a multi-qubit register. In the
hypercube representation, it projects the state onto one or other of
the $2D$ {\it faces} of the hypercube, rather than individual sites.


\begin{figure}
\centering (a)\\ \scalebox{0.45}{\includegraphics{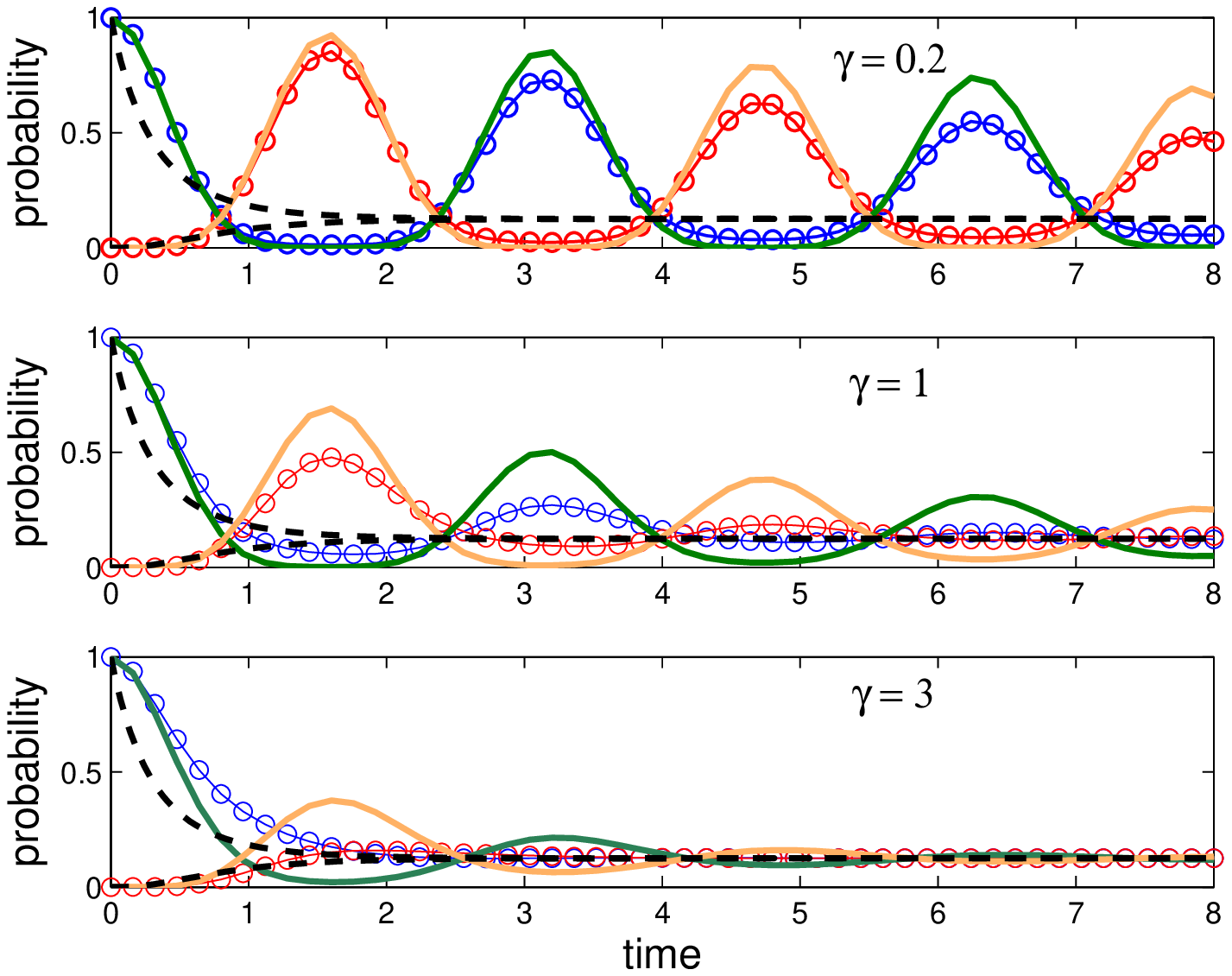}}\\ (b)\\ 
\scalebox{0.45}{\includegraphics{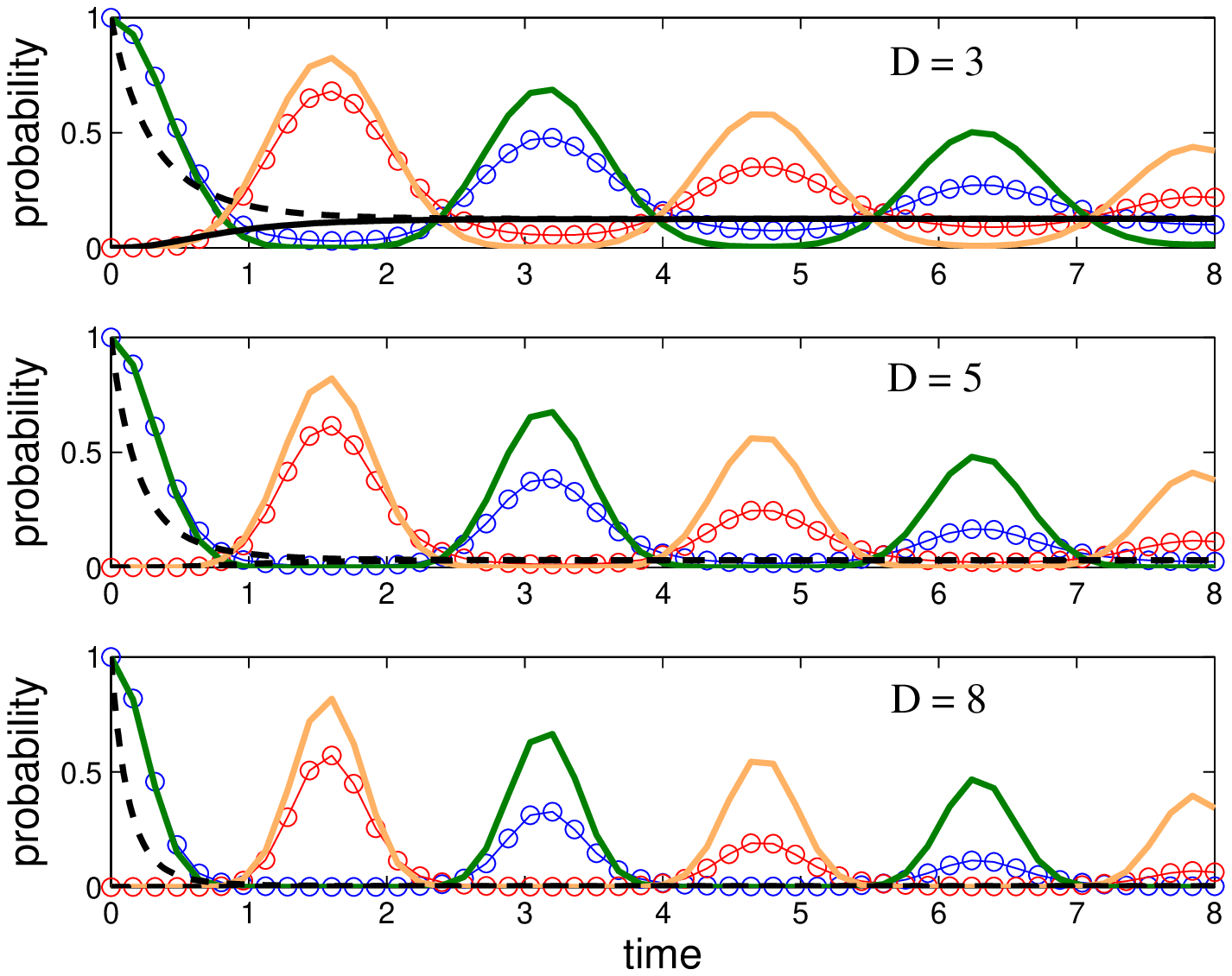}}
\caption{Comparison of the two decoherence mechanisms modelled by the master equation for the quantum walk on the hypercube. The walker in initially localized on a single corner. Plotted above in (a) and (b) is the probability that the walker remains at this location (blue and green), and the probability of being at the far corner (red and orange), over time: lines for qubit-based, circles for site-based decoherence. (a) shows results for $D=3$ with various $\gamma$, while (b) is for  $\gamma =0.5$ and differing $D$. In each plot, the dashed black line represents the classical behaviour.\label{fig::decoherence-comparison}}
\end{figure}

Let us consider the results for these 2 cases. For the case where
decoherence dephases individual qubits, the master equation is
\begin{equation}
\frac{d\rho(t)}{dt} = -i\left[H,\rho(t)\right] + \frac{\gamma}{2}
\sum_{k=1}^{D}\mathcal{D}\left[ \sigma^z_k\right]\rho(t),
\end{equation}
Since the qubits do not couple, the total density matrix is a
product of individual qubit density matrices, which we write in the
form
\begin{equation}
\rho(t) = \frac{1}{2} \left(\begin{array}{cc} 1 + z(t) & x(t)- i y (t) \\
x(t) + i y (t)  & 1 - z(t)  \end{array} \right),
\end{equation}
where $x(t)=\langle \sx(t) \rangle$, etc.

Then the solution is as follows. For the $x$-coordinate, we have
\begin{equation}
x(t) = x(0) e^{-\gamma t}.
\end{equation}
When $\gamma < 4\Delta_o$ the system is underdamped,resulting in
decaying oscillations around the origin in the $yz$-plane), given by
\begin{eqnarray}
y(t) & = & e^{-\gamma t/2}\left[ \frac{1}{\sqrt{1-r^2}}\left(z(0) -r
y(0)\right)\sin(\omega t) \right.\nonumber\\
&& \left.\quad\quad\quad\quad\quad+ y(0)\cos(\omega t)\frac{}{}\right],\\
z(t) &= & e^{-\gamma t/2}\left[\frac{}{}z(0) \cos(\omega t) \right. \nonumber\\
&& \quad\left.+ \frac{1}{\sqrt{1-r^2}}\left(r z(0) - y(0)
\right)\sin(\omega t)\right],
\end{eqnarray}
where $r=\gamma / 4 \Delta_o$ and $\omega = \sqrt{4\Delta_o^2 -
\gamma^2/2}$. In the overdamped regime, $\gamma>4\Delta$, we see
exponential decay,
\begin{eqnarray}
y(t) &=& \frac{-1}{2\sqrt{r^2-1}}\left[\mathcal{A}_+ e^{-\lambda_+
t}
+\mathcal{A}_- e^{-\lambda_- t}\right],\\
z(t) &=&\frac{-1}{2\sqrt{r^2-1}}\left[ \mathcal{B}_+ e^{-\lambda_+
t} +\mathcal{B}_- e^{-\lambda_- t}\right],
\end{eqnarray}
where $\lambda_{\pm} = 1/2 [\gamma \pm \sqrt{\gamma^2 -
16\Delta_o^2}]$, and
\begin{eqnarray*}
\mathcal{A}_{\pm} & = & \pm \left((r \mp \sqrt{r^2-1})y(0)-z(0)\right), \\
\mathcal{B}_{\pm} & = & \pm \left( y(0) - (r \pm \sqrt{r^2-1})z(0)\right).\\
\end{eqnarray*}

Suppose the walker is initialized at the origin (`$0$') node, so that
$z(0)=1$ and  $x(0)=y(0)=1$ for each qubit. The probability of being found at this
same site after a time $t$ is $P_{00}(t) =
\left(\frac{1+z(t)}{2}\right)^D$, while the probability of being at
the far corner is $P_{D0}(t) = \left(\frac{1-z(t)}{2}\right)^D$.

Consider now the results for the case where the master equation
projects onto each site. Figure \ref{fig::decoherence-comparison}
compare these `site'-based results (calculated numerically) with the
single-qubit based decoherence. The difference in the resulting
dynamics is easy to understand. In site-based decoherence, all
off-diagonal elements of the total density operator, which describe
quantum interference, decay at the same rate. However because
single-qubit dephasing monitors which \emph{face} of the hypercube
the walker is located on, different off-diagonal elements decay at
different rates; the element $\rho_{mn}$ decays at rate
$-(C/D)\gamma$, where $D$ is the number of qubits, and $1\leq C \leq
D$ is a constant, given by the Hamming weight of the difference
between the binary representations of the two indices $m$ and $n$.

An interesting question arising from these results is to what extent
the useful properties of a decohering quantum walk, with a
site-based environmental interaction \cite{KT03,Ric06}, can be
simulated in a multi-qubit representation. These issues will be
addressed in a forthcoming article \cite{DHS07} where a detailed
analysis of different decoherence channels acting on the quantum
walk on the hypercube, with a focus on those characteristics
important for potential algorithms, is presented.

\subsubsection{Hypercube coupled to an oscillator bath}

Let us now see what one gets by coupling the hypercube to a bath of
oscillators. We assume a Hamiltonian ${\cal H} = H_o^{HC} + V +
H_{osc}$, with $H_o^{HC}$ given in (\ref{Do}) above, and with
oscillator Hamiltonian given by
\begin{equation}
H_{osc} =  \sum_{q = 1}^{N_o} ( {p_q^2 \over m_q} + m_q \omega_q^2
x_q^2 )
 \label{H-osc}
\end{equation}
in terms of oscillators $\{ x_q \}$, and a diagonal interaction $V =
\sum_{q=1}^{N_o} v_n^{z}(q) \hat{\tau}_n^z  x_q$ between oscillators
and the qubits (note in line with the remarks above, this
interaction is {\it not} diagonal in the hypercube node basis).

Implicit but not mentioned in these Hamiltonians is a UV cutoff
$\Omega_o$; it is assumed that modes at energy scales $> \Omega_o$
do not couple to the walker. Suppose we now choose a lower UV cutoff
$\tilde{\Omega}_C$. Then the result is (i) a renormalisation of
$\Delta_o$, and (ii) an extra inter-qubit interaction term
$V_{nm}^{zz} \hat{\tau}_n^z \hat{\tau}_m^z$ in the effective
Hamiltonian, with
\begin{equation}
V_{nm}^{zz}(\tilde{\Omega}_C) \;=\;
\int_{\tilde{\Omega}_C}^{\Omega_o} {d\omega \over \pi}
{J_{nm}^{zz}(\omega) \over \omega}
 \label{delV}
\end{equation}
where the function $J_{nm}^{zz}(\omega)$ is defined in terms of the
propagator ${\cal D}({\bf R},\omega) = \sum_{\bf q} {\cal D} ({\bf
q},\omega) \exp[i {\bf q}\cdot{\bf R}]$ for the oscillator modes
(the zero-temperature propagator has the form ${\cal D} ({\bf
q},\omega) = 2\omega_{\bf q}/[\omega^2 - \omega_{\bf q}^2]$, with
the usual Matsubara generalisation to finite $T$). One has
\begin{equation}
J_{nm}^{zz}(\omega) \;=\; {\pi \over 2} \sum_{\bf q} {\vert
c_n^z({\bf q}) c_m^z({\bf q}) \vert \over \omega_{\bf q} }{\cal
D}({\bf R}_{nm},\omega_{\bf q}) \delta(\omega - \omega_{\bf q})
 \label{Jnm}
\end{equation}
where we assume qubits $n$ and $m$ separated by a radius vector
${\bf R}_{nm}$. The crucial point is that consistent calculations of
the qubit dynamics must give the same results for either the `bare'
form without interactions and with the bare $\Delta_o$, or the
lower-energy form with the renormalisation and the interaction.

To say anything more about this dynamics we need the form of the
bath spectrum and couplings - the details are very lengthy and will
be given elsewhere. But note that the dynamics is now utterly
different from what was found using the master equation approach -
the induced couplings between qubits generate entanglement between
them (in hypercube language they generate couplings across the
diagonals), as well as long-time tails in the single-qubit dynamics. This is seen already in solutions for the single qubit, ie., in the
well-known spin-boson model\cite{Wei99}, and in the solution for a
pair of qubits interacting with oscillators\cite{Pis98}.

\subsection{Hyperlattice Walker}
 \label{sec:hypL-SB}

The quantum walk on the $d$-dimensional hyperlattice, with
Hamiltonian (\ref{H-HLo}), is also trivially solvable. For a walker
initialized at the origin (centre) of the hyperlattice, $\vec{n} =
0$, the probability of being at some other site, with the position
described by the $d$-dimensional vector $\vec{n}$, after some time
$t$ is given by
\begin{equation}\label{eq::hyperlat}
P_{\vec{n}0}(t) = \prod_{\mu=1}^d J_{n_{\mu}}^2 (z),
\end{equation}
where $J_m(z)$ is the $m^{th}$ order Bessel function, of
dimensionless time $z=2\Delta t$. The purely quantum evolution is
characterized (at long times) by $P_{00}(t)\propto 1/t^d$ and by
mean-square displacement -- given by $\langle n^2(t) \rangle =
\sum_{\vec{n}} n^2 P_{\vec{n}0}(t)$ -- behaving as $\langle n^2(t)
\rangle \propto t^2$.

This result is often compared to classical diffusion on a
hyperlattice, where $P^{(cl)}_{00}(t)\propto 1/t^{d/2}$ and $\langle
n_{cl}^2(t) \rangle \propto t$.

\subsubsection{Master equation approach}

\begin{figure}[h!]
\centering
(a)\\
\scalebox{0.455}{\includegraphics{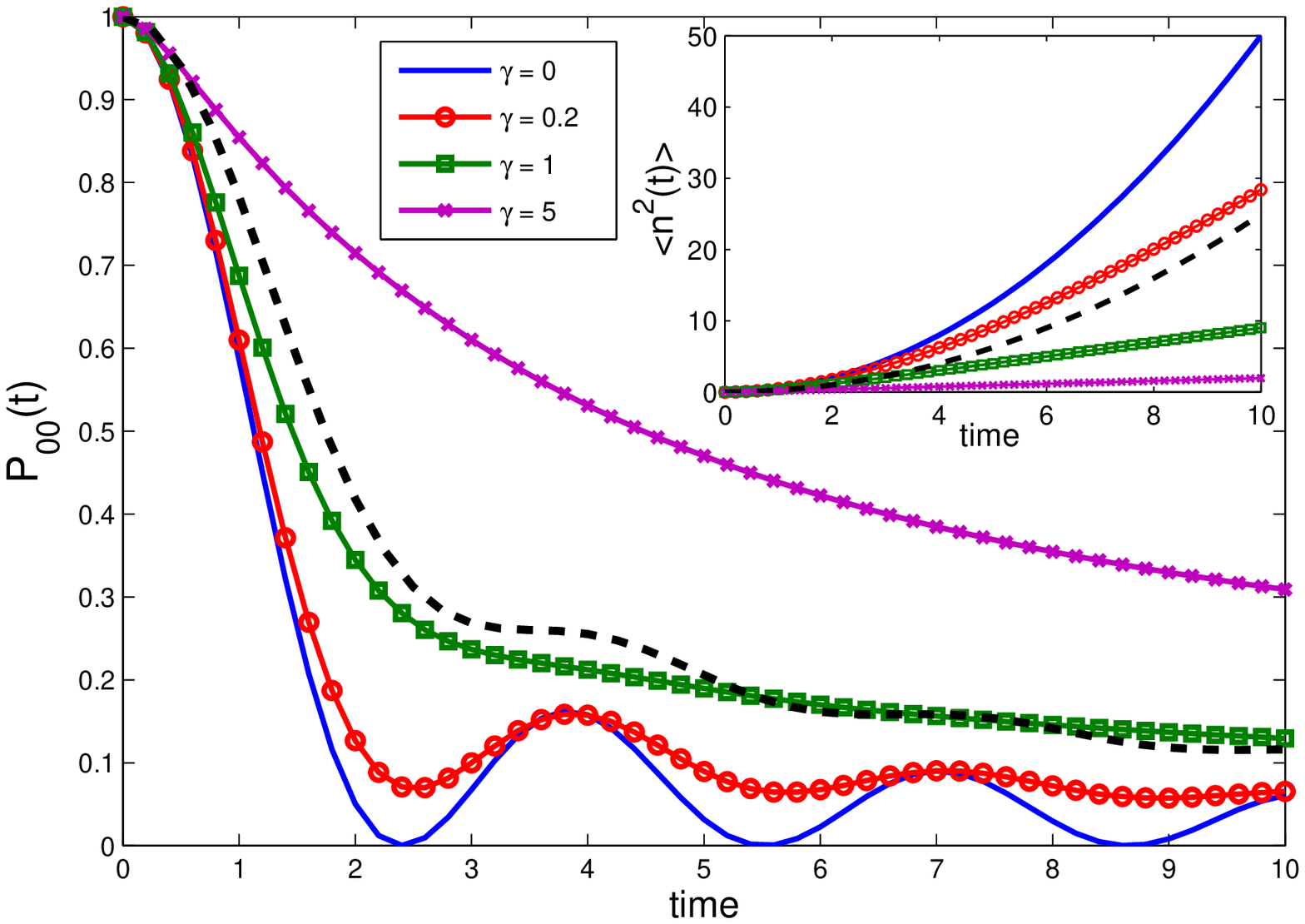}}\\
(b)\\
\scalebox{0.45}{\includegraphics{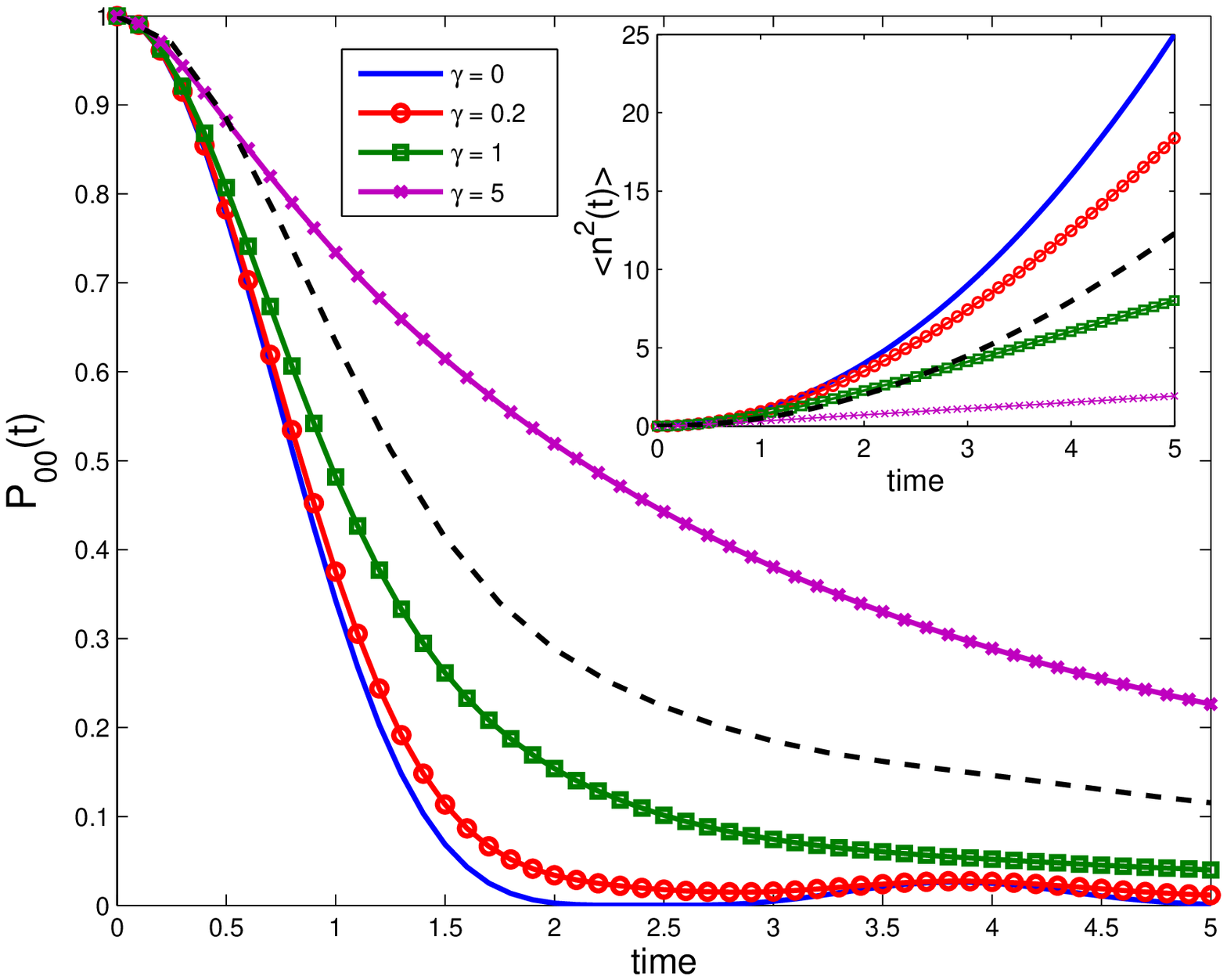}}\
\caption{Results for decoherence of the quantum walk on the \emph{hyperlattice} in (a) one dimensions and (b) two dimensions.
The coloured lines are for the dynamics described by the master
equation (\ref{eq::master}), with site
measurements, while the dashed black line is for the walker
strongly coupled to the spin-bath (\ref{eq::Psbath}). The
walker is initialized at the origin (centre of lattice) and
plotted above is the probability of remaining at the
origin over time, while inset is the mean-square displacement.
The differing behaviour for the spin-bath decoherence is clear
in the comparative behaviour of the probability to remain at
the origin and the mean-square displacement.} \label{fig::dec-HL}
\end{figure}

We first consider decoherence modelled by Eq. (\ref{eq::master}) with
\begin{equation}
\left\{M_{\vec{n}}\right\} = \left\{\ket{\vec{n}} \! \bra{\vec{n}}\right\}
\end{equation}
where $\vec{n}$ is the lattice vector defining a given site on the
hyperlattice, and we initialize the particle at the origin of the
lattice. This corresponds to an environment which monitors the
position of the walker -- an assumption consistent with a diagonal
(on-site) coupling to the environment. We consider both the
probability of the particle remaining at the origin, $P_{00}(t)$,
and the mean square displacement. Results from numerical
calculations, for $d=1,2$ are presented in figure \ref{fig::dec-HL}.
As $\gamma$ increases from zero, the quantum walk begins to behave
more classically, as exemplified in the behaviour of $\langle n^2(t)
\rangle$. However, as $\gamma$ becomes large, we see a quantum Zeno
effect, with the particle  trapped at the origin.

This transition to classical diffusion is very commonly found when
we couple the position of the walker to an environment - it is not
just found in memory-free noise models like this one, but also in
more realistic oscillator bath models\cite{CL83}. However it is by
no means generally valid. To see this, we consider what happens with
a spin bath environment.

\subsubsection{Hyperlattice coupled to a Spin Bath}

Consider a spin-bath of two-level systems which couples
\emph{non-diagonally} to the walker; the environment then monitors
the transitions between nodes. Suppose that each time the quantum walker hops between nodes it can
flip the $k^{th}$ bath spin, $\vec{\sigma}_k$, with complex flipping
amplitude $\alpha_k$. This gives an effective Hamiltonian,
\begin{equation}
H = \Delta_o \sum_{<ij>} \left\{ c_i^{\dagger} c_j\;
\cos\left(\sum_k \alpha_k \sigma_k^x \right)  + H.c. \right\}.
 \label{Htop}
\end{equation}
Hamiltonians of this form have previously been derived for a number
of physical systems\cite{PS00}. The strength of the decoherence is
parametrized by $\lambda = \sum_k \alpha_k^2$, which measures the
number of bath spins flipped during a transition between 2 nodes. We
let $\lambda \gg 1$, thereby considering the limit of strong
decoherence.

The dynamics of this model are exactly solvable \cite{PS06}, with
the reduced density operator of the walker given by a phase average
over the propagator of the `free' quantum walk. Taking the strong
decoherence limit, with the walker initially at the origin, we have
\begin{equation}\label{eq::Psbath}
P_{\vec{n}0}(t) = \int_0^{2\pi} \frac{d\varphi}{2\pi}
\prod_{\mu=1}^{d} J_{n_{\mu}}^2(z\cos\varphi).
\end{equation}
While for strong decoherence one might expect long-time diffusive
behaviour, what we actually see is quite different. For long times
($z \gg 1$), one has the limiting short-distance behaviour (return
probability) given by
\begin{equation}
P_{00}(z \to \infty ) \approx \frac{1}{\pi}\int_{-\infty}^{\infty}
d\varphi J_0^{2d}(z \varphi ) = \frac{A_d}{\Delta_o t } \;,
 \label{Pt0}
\end{equation}
where $A_d=(2\pi)^{-1}\:\int_{-\infty}^{\infty} dx J_0^{2d}(x)$ is a
constant (in $d=1$ there is an additional $\ln(2\Delta_o t)$
factor). This gives a divergence in the total time spent at the
origin; the interaction with the bath causes `sub-diffusive'
behaviour near the origin and a form of quasi-localization. However
quite remarkably, the mean-square displacement calculated from Eq.
(\ref{eq::Psbath}) is
\begin{equation}
 \langle n^2(t) \rangle = \frac{d}{2}\left(\Delta t\right)^2,
\end{equation}
which is but a factor of two smaller than the coherent quantum
evolution! The exact solution\cite{PS06} shows that the density
matrix has one component showing quasi-localization at the origin,
coexisting with another showing coherent ballistic dynamics far from
the origin. This is quite a departure from the classical diffusion
expected from a strong coupling to the environment! This behaviour
arises because the environmental coupling does not distinguish
different walker positions on the hyperlattice, or the direction of
transitions between nodes. It only records when a transition between
nodes has occurred, allowing for constructive interference over many
differing paths on the graph. For more
details see Ref. \cite{PS06}.\\

We emphasize here that the graph over which the quantum walk takes
place can be represented using different sets of basis states; there
is no general principle forcing environmental couplings to
distinguish different nodes or transition directions in these
different encodings. In the design of quantum computers and certain
search algorithms, the above result shows the importance of
investigating quantum walks for which environmental couplings do not
distinguish different `position' (or `nodal') states in the
information space in which the quantum walk is encoded \cite{DHS07}.

\section{Concluding Remarks}

The extent to which quantum information processing and entanglement
are robust to decoherence is a central question in quantum
computation, and by implication in quantum walk theory. The two
examples we have discussed here highlight the importance of the
choice of `information space' in which the structure of the walk is
encoded. To fully appreciate the potential impact for quantum walk
algorithms, one must focus on the relevant algorithmic properties of
the quantum walk, and the effect of different decoherence processes.

Quantum walks were originally devised as a new way to develop
quantum algorithms. However we have found that they may actually be
more useful in understanding the effect of the environment on
quantum information processing, using mappings between arbitrary
quantum information systems and quantum walk systems. Most of the
promising architectures for quantum computing, especially
solid-state proposals, have qubits which couple to quantum
environments which cannot be treated simply as `noise' sources. In
reality the coupling to the environment generates complex
correlated errors, as well as highly non-local (in space and time)
correlations in the decoherence dynamics. By representing this
complex dynamics in terms of quantum walks coupled to general
quantum environments, one makes it possible to arrive at very useful
results.

All of these investigations may usefully be understood as a new
direction in research on the quantum dynamics of various systems
(moving particles, spins, etc.) on different kinds of lattice and
graphs. This field has an interesting history, with many
applications in statistical mechanics \cite{StM}, and the theory of
disordered systems \cite{disor}. It seems likely that methods
imported from these fields will be very useful in understanding
quantum information processing.\\


We would like to thank NSERC, PITP and PIMS for support.

\end{document}